\documentclass{article}
\usepackage{amsmath,amssymb,verbatim,url}

\def\N{\mathbb{N}}
\def\Z{\mathbb{Z}}

\def\K{\mathbb{K}}

\newcommand {\dd}{\,{\rm div}\,}

\newcommand {\ord}{\mathop{\rm ord}}

\newcommand {\gf}{g}
\newcommand {\nested}{\text{\large\rm NS}}
\newcommand{\proof}{\noindent{\it Proof:\/ }}
\newcommand{\qed}{\hfill $\square$}

\newtheorem{theorem}{\bf Theorem}
\newtheorem{corollary}{\bf Corollary}
\newtheorem{lemma}{\bf Lemma}
\newtheorem{proposition}{\bf Proposition}

\newtheorem{definition}{\bf Definition}
\newtheorem{example}{\bf Example}

\begin{document}

\title{Convolutions of Liouvillian Sequences}

\author{Sergei A.~Abramov\footnote{Dorodnicyn Computing Center,  Federal Research Center ``Computer Science and Control'', Russian Academy of Sciences, Moscow, Russia} \and
Marko Petkov\v sek\footnote{Faculty of Mathematics and Physics, University of Ljubljana, and Institute of Mathematics, Physics and Mechanics, Ljubljana, Slovenia}
\and Helena Zakraj\v sek\footnote{Faculty of Mechanical Engineering, University of Ljubljana, Slovenia}}

\date{}

\maketitle

\begin{abstract}
While Liouvillian sequences are closed under many operations, simple examples show that they are not closed under convolution, and the same goes for d'Alembertian sequences. Nevertheless, we show that d'Alem\-ber\-ti\-an sequences are closed under convolution with \emph{rationally} d'Alembertian sequences, and that Liouvillian sequences are closed under convolution with \emph{rationally} Liouvillian sequences.

\end{abstract}

\textbf{Keywords:}
 (rationally) d'Alembertian sequences; (rationally) Liouvillian sequences; closure properties; convolution

\medskip
MSC (2010) 68W30; 33F10

\section{Introduction}

Let $\K$ be an algebraically closed field of characteristic 0, $\N$ the set of nonnegative integers, and $\K^\N$ the set of all sequences with terms in $\K$. 

\begin{definition}
\label{poly(q)rat}
A sequence $\langle a_n\rangle_{n=0}^\infty \in \K^\N$ is:
\begin{itemize}
\item {\em polynomial\/} if there is $p \in \K[x]$ such that $a_n = p(n)$ for all $n \in \N$,
\item {\em rational\/}  if there is $r \in \K(x)$ such that $a_n = r(n)$ for all large enough $n$,
\item {\em quasi-rational (}cf.\ {\em\cite{Abr91})\/} if there are $d \in \N$, rational functions $r_1, r_2, \ldots, r_{d} \in \K(x)^*$, and $\alpha_1, \alpha_2, \ldots, \alpha_{d} \in \K^*$ such that $a_n = \sum_{i=1}^{d} r_i(n)\alpha_i^n$  for all large enough $n$.
\end{itemize}
\end{definition}
\begin{definition}
A sequence $\langle a_n\rangle_{n=0}^\infty \in \K^\N$ is
{\em $P$-recursive} or {\em  holonomic} if there are $d\in\N$ 
and polynomials $p_0, p_1,\ldots, p_d \in \K[n]$, $p_d \ne 0$, such that
\[
p_d(n) a_{n+d} + p_{d-1}(n) a_{n+d-1} + \cdots + p_0(n) a_{n} \ =\ 0
\]
for all $n\in \N$. In particular, a holonomic sequence is {\em  hypergeometric} if
\begin{enumerate}
\item there are $p, q\in \K[n] \setminus\{0\}$ such that
\[
q(n)\, a_{n+1} + p(n) \, a_n \ =\ 0 \ \ {\rm for\ all\ } n \ge 0,
\]
\item there is an $N \in \N$ such that $a_n \ne 0$ for all $n \ge N$.
\end{enumerate}
The set of all holonomic sequences in $\K^\N$ will be denoted by ${\cal P}(\K)$, and the set of all hypergeometric sequences in $\K^\N$ by ${\cal H}(\K)$.
\end{definition}

\begin{example}
Some hypergeometric sequences:
\medskip
\begin{itemize}
\item $a_n = c^n$ \ where $c \in \K^*$ {\em (geometric sequences)}
\item $a_n = p(n)$ where $p \in \K[n]\setminus\{0\}$ {\em (nonzero polynomial sequences)}
\item $a_n = r(n)$ for all large enough $n$\ where $r \in \K(n)^*$ {\em (nonzero rational sequences)}
\item $a_n = n!$
\item $a_n = \displaystyle\binom{2n}{n}$
\end{itemize}
\end{example}

Our long-term goal is to design algorithms for finding {\em explicit representations} of holonomic sequences in terms of
\begin{itemize}
\item some basic sequences expressible in closed form (such as, e.g., rational sequences or hypergeometric sequences),
\item some common operations with sequences which preserve holonomicity.
\end{itemize}
Here we are particularly interested in those explicit representations whose admissible operations include convolution, also known as Cauchy product.

\medskip
\textit{A short overview of the paper:} In Section 2 we introduce several well-known holonomicity-preserving operations (see also \cite{Pet06}), and call an operation $\Omega$ \emph{local} if the equivalence relation holding between sequences which eventually agree is a congruence w.r.t.\ $\Omega$ (i.e., if, when given equivalent operands, $\Omega$ yields equivalent results). Convolution is not local which makes working with it a little harder. In line with the above-described scheme of defining classes of explicit representations, we describe in Section 3 the rings of d'Alembertian, Liouvillian, rationally d'Alembertian (cf.\,\cite{Abr93}), and rationally Liouvillian sequences. In the former two cases, the basis is the set of hypergeometric sequences, and in the latter two, the ring of rational sequences. While the former two are not closed under convolution, we show in Sections 4 resp.\ 5 that the convolution of a d'Alembertian sequence with a (quasi-)rationally d'Alembertian sequence (see Def.\ \ref{qrdalemb}) is d'Alembertian (Corollary \ref{dalemb*ratdalemb}), and the convolution of a Liouvillian sequence with a (quasi-)rationally Liouvillian sequence (see Def.\ \ref{qrliouv}) is Liouvillian (Corollary  \ref{liou*rat}). We divide the proof (for d'Alembertian sequences) into two parts: Theorem \ref{main} deals with the ``ideal'' case where the minimal annihilators of the hypergeometric resp.\ rational sequences in the two factors are nonsingular, and Corollary \ref{dalemb*ratdalemb} takes care of the rest. In Section 6 we list some open problems and present an algorithm for finding solutions of a linear recurrence that are convolutions with a given hyperexponential sequence.

\medskip
\textit{Some further definitions and notation:}
\begin{itemize}
\item For $x \in \K$ and $n \in \N$, we denote by $x^{\underline{n}} := \prod_{j=0}^{n-1} (x-j)$ the $n$-th falling power of $x$.
\item For $n,m\in\N$, $m\ge 1$, we denote by $n \dd m := \lfloor\frac{n}{m}\rfloor$ the quotient, and by $n \bmod m := n - m \lfloor\frac{n}{m}\rfloor$ the remainder in integer division of $n$ by $m$.
\item The \emph{shift operator} $E:\ \K^\N \to \K^\N$ is defined for all $a \in \K^\N$, $n \in \N$ by $E(a)_n\ =\ a_{n+1}$, and for $k \in \N$, its $k$-fold composition with itself is denoted by $E^k$. For $d \in \N$ and $p_0, p_1, \ldots, p_d \in \K[n]$ such that $p_d \ne 0$, the map $L =  \sum_{k=0}^d p_k(n)\,E^k:\ \K^\N \to \K^\N$ is a \emph{linear recurrence operator} of order $\ord L = d$ with polynomial coefficients. We denote the Ore algebra of all such operators (with composition as multiplication) by $K[n]\langle E \rangle$.
\end{itemize}

\section{Operations with sequences}

\begin{theorem}
\label{ops}
${\cal P}(\K)$ is closed under the following operations:
\begin{itemize}
\item unary operations $a \mapsto c$
\begin{enumerate}
\item {\em scalar multiplication:} \quad $c_n = \lambda a_n$\ \,where\ $\lambda \in \K$
\item {\em shift:} \quad $c_n = E(a)_n = a_{n+1}$ 
\item {\em inverse shift:} \quad $c_n = E_{\lambda}^{-1}(a)_n = \left\{
\begin{array}{ll}
 a_{n-1}, & n \ge 1 \\
\lambda, & n = 0
\end{array}
\right.$\ \,where\ $\lambda \in \K$
\item {\em difference:} \quad $c_n = \Delta a_n = a_{n+1} - a_n$ 
\item {\em partial summation:} \quad $c_n = \sum_{k=0}^n \,a_k$ 
\item {\em multisection:} \quad $c_n = a_{mn+r}$\ \,where\ $m \in \N\setminus\{0\}, \ r \in \{0,1, \dots, m-1\}$
\phantom{aaaaaaaaaaaaaaaaa}\, \emph{(the $r$-th $m$-section of $a$)}
\end{enumerate}
\item binary operations $(a,b) \mapsto c$
\begin{enumerate}
\setcounter{enumi}{6}
\item {\em addition:} \quad $c_n = a_n + b_n$
\item {\em multiplication:} \quad $c_n = a_n b_n$ 
\item {\em convolution:} \quad $c_n = (a * b)_n = \sum_{k=0}^n \,a_k b_{n-k}$ 
\end{enumerate}
\item polyadic operations $(a^{(0)}, a^{(1)}, \ldots, a^{(m-1)}) \mapsto c$ \ where $m \in \N \setminus \{0\}$
\begin{enumerate}
\setcounter{enumi}{9}
\item {\em interlacing:} \quad $c_n\ =\ \Lambda(a^{(0)}, a^{(1)}, \ldots, a^{(m-1)})_n \ =\  \left(\Lambda_{j=0}^{m-1}a^{(j)}\right)_n$ \\
\phantom{aaaaaaaaaaaaaaa} $ =\ a^{(n\bmod m)}_{n\dd m}$
\end{enumerate}
\end{itemize}
\end{theorem}

\proof  Let  $L =  \sum_{k=0}^d p_k(n)\,E^k \in \K[n]\langle E \rangle \setminus \{0\}$ be such that $L(a) = 0$.
\begin{enumerate}
\item $L(\lambda a) = \lambda L(a) = 0$, so $\lambda a \in {\cal P}(\K)$.
\item Let $L' := \sum_{k=0}^d p_k(n+1)\,E^k \in \K[n]\langle E \rangle \setminus \{0\}$. Then
\[
L'(E(a)) = \left(\sum_{k=0}^d p_k(n+1)\,E^{k+1}\right)(a) = (EL)(a) = E(L(a)) = 0,
\]
so $E(a) \in {\cal P}(\K)$.
\item Note that $E E_{\lambda}^{-1} = {\rm id}_{\K^\N}$, hence
\[
(LE)(E_{\lambda}^{-1}(a)) = (L(E E_{\lambda}^{-1}))(a) = L(a) = 0,
\]
so $E_{\lambda}^{-1}(a) \in {\cal P}(\K)$.
\item This follows from items 2, 1, and 7.
\end{enumerate}
For proofs in the remaining six cases, see \cite{HS, Stanley80, Stanley99}. 
\qed

Note that operations 7 -- 9 are associative (and commutative), hence they can also be considered as polyadic.

\begin{definition} {\em \cite{Stanley80}}
Sequences $a, b \in \K^\N$ are {\em equivalent} if there is an $N \in \N$ s.t.
\[
a_n = b_n \ \ {\rm for\ all\ } n \ge N
\]
or equivalently, s.t.\ $E^N(a) = E^N(b)$. We denote this relation by\ $\,\sim$, and call its equivalence classes {\em germs (at $\infty$ of functions $\N \to \K$)}.

We say that a set of sequences ${\cal C} \subseteq \K^{\N}$ is \emph{closed under equivalence} if $a \in {\cal C}$ and $a \sim a'$ implies $a' \in {\cal C}$.
\end{definition}

\begin{proposition}
The set ${\cal H}(\K)$ is closed under equivalence.
\end{proposition}

\proof Assume that $a \in {\cal H}(\K)$ and $a' \sim a$. Then there are $p, q \in \K[n] \setminus \{0\}$ and $N \in \N$ s.t.\ $q(n) a_{n+1} + p(n) a_n = 0$ for all $n \in \N$, and $a_n' = a_n \ne 0$ for all $n \ge N$. Hence 
\[
n^{\underline N}\, q(n) a_{n+1}' + n^{\underline N}\, p(n) a_n' = 0
\]
for all $n \in \N$, so $a' \in {\cal H}(\K)$.
\qed

\begin{proposition}
\label{closedeq}
Let ${\cal C} \subseteq \K^{\N}$ be a class of sequences closed under all inverse shifts and addition, and such that $0 \in {\cal C}$. Then $\cal C$ is closed under equivalence. 
\end{proposition}

\proof   Let  $a \in {\cal C}$ and $a' \sim a$. Then there are $k \in \N$ and $\lambda_0, \lambda_1, \ldots, \lambda_k \in \K$ s.t.
\[
a' - a = \langle \lambda_0, \lambda_1, \ldots, \lambda_k, 0,0,0,\ldots \rangle = E_{\lambda_0}^{-1} E_{\lambda_1}^{-1} \cdots E_{\lambda_k}^{-1} (0),
\]
so $a' = a + E_{\lambda_0}^{-1} E_{\lambda_1}^{-1} \cdots E_{\lambda_k}^{-1} (0) \in {\cal C}$. 
\qed

\begin{corollary}
The holonomic ring ${\cal P}(\K)$ is closed under equivalence.
\end{corollary}

\begin{definition}
An operation $\omega$ on $\K^\N$ is {\em local} if $\sim$ is a congruence w.r.t.\ $\omega$.
\end{definition}

\begin{proposition} 
\smallskip
The following operations are local: scalar multiplication, shift, inverse shift, difference, multisection, addition, multiplication, interlacing.
\end{proposition}

\proof   Straightforward. 
\qed

\begin{example} 
Partial summation is not local:  Let, e.g.,
\begin{eqnarray*}
a &=& \langle 0,0,0,\ldots\rangle, \\
b &=& \langle 1,0,0,\ldots\rangle.
\end{eqnarray*}
Then $a \sim b$ but
\[
\sum_{k=0}^n a_k = 0\ \ \not\sim\ \ \sum_{k=0}^n b_k = 1.
\]
Since $\sum_{k=0}^n \,a_k = (a * 1)_n$, it follows that convolution is not local either.
\end{example}

When dealing with local operations, it is customary to work with germs of sequences which simplifies the statements of results and their corresponding proofs. Since here we are especially interested in the non-local operations of convolution and partial summation, we have to work with sequences themselves. In this situation, the following auxiliary results are useful.

\begin{lemma} 
\label{conveq}
Let $a, b, \varepsilon, \eta \in \K^\N$ with $\varepsilon, \eta \sim 0$. Then:
\begin{itemize}
\item[\rm (i)] $a\, \varepsilon \sim 0$,
\item[\rm (ii)] $\sum_{k=0}^n \varepsilon_k \sim C$ for some $C \in \K$,
\item[\rm (iii)] $a * \varepsilon = \sum_{k=0}^{N} \varepsilon_k E_0^{-k} (a)$ for some $N \in \N$,
\item[\rm (iv)] $\varepsilon * \eta \sim 0$,
\item[\rm (v)] $(a+\eta) * (b+\varepsilon)\ \sim\ a * b\ + \sum_{i=0}^{N_1} \varepsilon_i E_0^{-i} (a)\ + \sum_{j=0}^{N_2} \eta_j E_0^{-j} (b)$ \\[2pt]
for some $N_1, N_2 \in \N$.

\end{itemize}
\end{lemma}

\noindent
{\em Proof:\,} 
\begin{itemize}
\item[\rm (i)] This follows from locality of multiplication. 
\item[\rm (ii)] Let $N \in \N$ be such that $\varepsilon _k = 0$ for $k > N$. Write  $C = \sum_{k=0}^N \varepsilon_k$. For $n \ge N$ we have $\sum_{k=0}^n \varepsilon_k = \sum_{k=0}^N \varepsilon_k$, so $\sum_{k=0}^n \varepsilon_k \sim C$.
\item[\rm (iii)] Let $N \in \N$ be such that $\varepsilon_k = 0$ for $k > N$. Then for all $n \in \N$,
\begin{equation}
\label{suma}
(a * \varepsilon)_n = \sum_{k=0}^n \varepsilon_k a_{n-k} = \sum_{k=0}^{\min\{n,N\}} \varepsilon_k  E_0^{-k} (a)_n = 
\bigg(\sum_{k=0}^{N} \varepsilon_k E_0^{-k} (a)\bigg)_n
\end{equation}
where the last equality follows from the fact that $E_0^{-k} (a)_n = 0$ for $k > n$.
\item[\rm (iv)] Let $N_1, N_2 \in \N$ be such that $\varepsilon_i = 0$ for $i > N_1$ and $\eta_j = 0$ for $j > N_2$. Assume that $n > N_1 + N_2$. Then $k > N_1$ or $n-k > N_2$ for every $k \in \N$, therefore
\[
(\varepsilon * \eta)_n = \sum_{k=0}^n \varepsilon_k \eta_{n-k} = 0
\]
for all such $n$, so $\varepsilon * \eta \sim 0$.
\item[\rm (v)] By bilinearity and commutativity of convolution we have
\[
(a+\eta) * (b+\varepsilon) = a * b + a * \varepsilon + b * \eta + \varepsilon * \eta.
\]
The claim now follows from (iii) and (iv).
 \qed
\end{itemize}

\section{Some classes of explicitly representable holonomic sequences}

Here we list some subrings of the holonomic ring $({\cal P}(\K),+,\cdot)$, defined as closures of a basic set of sequences under a set of holonomicity-preserving operations.
\begin{definition} 
The ring of {\em d'Alembertian sequences} ${\cal A}(\K)$ is the least subring of $({\cal P}(\K),+,\cdot)$ which contains ${\cal H}(\K)$ and is closed under
\begin{itemize}
\item shift,
\item all inverse shifts,
\item partial summation.
\end{itemize}
\end{definition} 

\begin{example}
Some d'Alembertian sequences:
\medskip
\begin{itemize}
\item derangement numbers $d_n\ =\ n! \sum_{k=0}^n \frac{(-1)^k}{k!}$
\item harmonic numbers $H_n\ =\ \sum_{k=1}^n \frac{1}{k}\ =\ \sum_{k=0}^n \frac{1}{k+1} - \frac{1}{n+1}$
\end{itemize}
\end{example}

\begin{definition}
For $d \in \N \setminus \{0\}$ and $a^{(1)},  a^{(2)},  \ldots, a^{(d)} \in \K^\N$, we shall denote by
\[
\nested\left(a^{(1)},  a^{(2)},  \ldots, a^{(d)} \right)\ = \ \nested_{i=1}^d a^{(i)}
\]
the sequence $a \in \K^\N$ defined for all $k_1 \in \N$ by
\begin{equation}
\label{dalemb0}
a_{k_1}\ :=\ \left(\nested_{i=1}^d a^{(i)}\right)_{k_1} \ =\  a^{(1)}_{k_1} \sum_{k_2=0}^{k_1} a^{(2)}_{k_2} \sum_{k_3=0}^{k_2} a^{(3)}_{k_3} \cdots \sum_{k_{d}=0}^{k_{d-1}} a^{(d)}_{k_d}
\end{equation}
and call it the \emph{nested sum} of sequences $a^{(1)},  a^{(2)},  \ldots, a^{(d)}$. We will call the number $d$ the \emph{nesting depth} of this particular representation of $a$.
\end{definition}

\begin{theorem}
\label{dalembprop}
Let $a \in \K^\N$. Then:
\begin{enumerate}
\item[\rm (i)] $a$ is d'Alembertian iff it can be written as a $\K$-linear combination (possibly empty) of nested sums of the form (\ref{dalemb0}) where $a^{(1)},  a^{(2)}, \ldots, a^{(d)} \in {\cal H}(\K)$,
\item[\rm (ii)] $a$ is d'Alembertian iff there are $d\in\N \setminus \{0\}$ and $L_1, L_2, \ldots,L_d  \in \K[n]\langle E\rangle$, each of order 1, such that $L_1 L_2 \cdots L_d(a) = 0$.
\end{enumerate}
\end{theorem}
For a proof, see \cite{AP94} or \cite{PZ13}. 

\begin{corollary}
\label{dalembRHS}
If $y \in \K^\N$ satisfies $L(y) = a$ where $L$ is a product of first-order operators and $a \in {\cal A}(\K)$, then $y \in {\cal A}(\K)$.
\end{corollary}

\proof
By Theorem \ref{dalembprop}.(ii), there are $d\in\N \setminus \{0\}$ and $L_1, L_2, \ldots,L_d  \in \K[n]\langle E\rangle$, each of order 1, such that $L_1 L_2 \cdots L_d(a) = 0$. Hence
\[
L_1 L_2 \cdots L_dL(y)\ =\ L_1 L_2 \cdots L_d(a)\ =\ 0,
\]
so, again by Theorem \ref{dalembprop}.(ii), $y \in {\cal A}(\K)$.
\qed

\begin{example}
It is straightforward to verify that for
\begin{itemize}
\item derangement numbers: $(E + 1)(E - (n + 1)) (d)\ =\ 0$,
\item harmonic numbers: $((n + 2)E - (n + 1))(E - 1) (H)\ =\ 0$.
\end{itemize}
\end{example}

\begin{definition} 
\label{liouv}
The ring of {\em Liouvillian sequences} ${\cal L}(\K)$ is the least subring of $({\cal P}(\K),+,\cdot)$ which contains ${\cal H}(\K)$ and is closed under
\begin{itemize}
\item shift,
\item all inverse shifts,
\item partial summation,
\item interlacing.
\end{itemize}
\end{definition} 

\begin{example}
Perhaps the simplest element of ${\cal L}(\K) \setminus {\cal A}(\K)$ is $a_n = n!!$, defined recursively by $a_0 = a_1 = 1$, $a_n = n a_{n-2}$ for $n \ge 2$. From
\[
n!! \ =\ \left\{
\begin{array}{ll}
2^k k!, & n = 2k, \\[2pt]
\frac{(2k+1)!}{2^k k!}, & n = 2k+1
\end{array}\right.
\]
we see that $n!!$ is the interlacing of two hypergeometric sequences, hence it is Liouvillian. On the other hand, its annihilating operator $L = E^2 - (n+2)$ has no nonzero d'Alembertian elements in its kernel.
\end{example}

\begin{theorem}
\label{interdAlemb}
A sequence $a \in \K^\N$ is Liouvillian if and only if it is an interlacing of d'Alembertian sequences.
\end{theorem}
For a proof, see \cite{Reu12} or \cite{PZ13}.

\begin{proposition}
\label{prop:equiv}
${\cal A}(\K)$ and ${\cal L}(\K)$ are closed under equivalence.
\end{proposition}

\noindent
{\em Proof:\,} This follows immediately from Proposition \ref{closedeq}. \qed

The ring of Liouvillian sequences ${\cal L}(\K)$ is, by Definition \ref{liouv} or by its immediate consequences, closed under all the operations listed in Theorem \ref{ops}, with possible exception of multisection and convolution. It is easy to see that both ${\cal A}(\K)$ and ${\cal L}(\K)$ are closed under multisection (cf.\ \cite{PZ13} and \cite{HS}). With convolution, the situation is much more varied already for hypergeometric operands, as demonstrated by the following three examples. To avoid having to name every sequence that we encounter, we will often use $a_n * b_n$ to denote either $(a * b)_n$ or $a * b$, with the precise meaning determined by the context.

\begin{example}
The convolution of $1/n!$ with itself
\[
\frac{1}{n!} * \frac{1}{n!}\ =\ \sum_{k=0}^n \frac{1}{k!(n-k)!}\ =\ \frac{1}{n!}\sum_{k=0}^n \binom{n}{k}\ =\ \frac{2^n}{n!}
\]
is hypergeometric.
\end{example}
  
\begin{example}
Zeilberger's Creative Telescoping algorithm {\em \cite{Zeil90, Zeil91}} shows that the convolution of $n!$ with itself
\[
y_n\ :=\ n! * n!\ =\ \sum_{k=0}^n k!(n-k)!
\]
satisfies the recurrence
\begin{equation}
\label{inhom}
2 y_{n+1} - (n+2)y_n\ =\ 2 (n+1)!
\end{equation}
which, together with the initial condition $y_0 = 1$, implies that
\[
y_n\ =\ \frac{(n + 1)!}{2^n} \sum_{k=0}^n \frac{2^k}{k + 1}.
\] 
This is a d'Alembertian sequence which is not hypergeometric, as shown by Gosper's summation algorithm {\em \cite{Gos78}}, or by algorithm Hyper {\em \cite{Pet92}}  applied to the homogenization $2y_{n+2}-(3n+7)y_{n+1}+(n+2)^2y_n = 0$ of {\em (\ref{inhom})}.
\end{example}

\begin{example}
\label{nonL}
Zeilberger's algorithm shows that the convolution of $n!$ with $1/n!$
\[
y_n\ :=\ n! * \frac{1}{n!}\ =\ \sum_{k=0}^n \frac{k!}{(n-k)!}
\]
satisfies the recurrence
\[
y_{n+2} - (n+2) y_{n+1} + y_n = \frac{1}{(n+2)!}
\]
which can be homogenized (by applying the annihilator $(n+3)E - 1$ of the right-hand side  to both sides) to $L y = 0$ where
\begin{equation}
\label{CLrec}
L = (n+3) E^3 - ( n^2 + 6 n + 10) E^2 + (2 n+5) E - 1.
\end{equation}
This recurrence has no nonzero Liouvillian solutions, as shown by the Hendriks-Singer algorithm {\em \cite{HS}}. So the convolution of hypergeometric sequences $n!$ and $1/n!$ is not Liouvillian.
\end{example}

As shown by Example \ref{nonL}, neither ${\cal A}(\K)$ nor ${\cal L}(\K)$ is closed under convolution. To obtain positive results, we define some further subrings of these rings by replacing hypergeometric sequences with (quasi-)rational sequences as their basis.

\begin{definition} 
\label{qrdalemb}
The ring of {\em (quasi-)rationally d'Alembertian sequences} ${\cal A}_{(q)rat}(\K)$ is the least subring of $({\cal P}(\K),+,\cdot)$ which contains all (quasi-)rational sequences over $\K$ and is closed under
\begin{itemize}
\item shift,
\item all inverse shifts,
\item partial summation.
\end{itemize}
\end{definition} 

\begin{example}
Harmonic numbers $H_n = \sum_{k=1}^n \frac{1}{k}$ are rationally d'Alembertian.
\end{example}

\begin{theorem}
\label{ratdalembprop}
A sequence $a \in \K^\N$ is (quasi-)rationally d'Alembertian iff it can be written as a $\K$-linear combination (possibly empty) of nested sums of the form (\ref{dalemb0}) where $a^{(1)},  a^{(2)}, \ldots, a^{(d)}$  are (quasi-)rational sequences.

\end{theorem}
The proof is analogous to that of Theorem \ref{dalembprop}(i). 

\begin{definition} 
\label{qrliouv}
The ring of {\em (quasi-)rationally Liouvillian sequences} ${\cal L}_{(q)rat}(\K)$ is the least subring of $({\cal P}(\K),+,\cdot)$ which contains all (quasi-)rational sequences over $\K$ and is closed under
\begin{itemize}
\item shift,
\item all inverse shifts,
\item partial summation,
\item interlacing.
\end{itemize}
\end{definition} 

\begin{example}
The interlacing of harmonic numbers $H_n = \sum_{k=1}^n \frac{1}{k}$ with generalized harmonic numbers of order 2, $H_n^{(2)} = \sum_{k=1}^n \frac{1}{k^2}$, is rationally Liouvillian.
\end{example}

\section{D'Alembertian sequences under convolution}
\label{dal}

\begin{proposition}
\label{dalembEa*Eb}
For all $k \in \N$ and all $a, b \in {\cal A}(\K)$, we have
\begin{equation}
\label{Ea*Ebeq}
a * b \in {\cal A}(\K)\ \Longleftrightarrow\ E^k (a) * E^k (b) \in {\cal A}(\K).
\end{equation}
\end{proposition}

\proof  Note that, for all $n \in \N$,
\begin{align*}
E^2(a * b)_n &= \sum_{k=0}^{n+2} a_k b_{n+2-k} =  a_0 b_{n+2} + a_{n+2} b_0 + \sum_{k=1}^{n+1} a_k b_{n+2-k} \\
&=  a_0 b_{n+2} + a_{n+2} b_0 + \sum_{k=0}^n a_{k+1} b_{n+1-k} \\
&= a_0 E^2(b)_n +  b_0 E^2(a)_n + (E(a) * E(b))_n.
\end{align*}
As ${\cal A}(\K)$ is closed under shift, scalar multiplication and addition, this implies 
\[
E^2(a * b) \in {\cal A}(\K)\ \Longleftrightarrow\ E(a) * E(b) \in {\cal A}(\K).
\]
By the closure of ${\cal A}(\K)$ under shift and all inverse shifts, we have
\[
a * b \in {\cal A}(\K)\ \Longleftrightarrow\ E^2(a * b) \in {\cal A}(\K),
\]
so
\begin{equation}
\label{a*b_Ea*Eb}
a * b \in {\cal A}(\K)\ \Longleftrightarrow\ E(a) * E(b) \in {\cal A}(\K).
\end{equation}
As ${\cal A}(\K)$ is closed under shift, we can replace $a$ by $E^k(a)$ and $b$ by $E^k(b)$ in (\ref{a*b_Ea*Eb}) and obtain
\[
E^k(a) * E^k(b) \in {\cal A}(\K)\ \Longleftrightarrow\ E^{k+1}(a) * E^{k+1}(b) \in {\cal A}(\K)
\]
for all $k \in \N$. Now (\ref{Ea*Ebeq}) follows by induction on $k$.
\qed

\begin{lemma}
\label{lemmaZ}
Let $d \in \N$, $a^{(1)}, a^{(2)}, \ldots, a^{(d)} \in \K^\N$, $\varepsilon^{(1)}, \varepsilon^{(2)}, \ldots, \varepsilon^{(d)} \in \K^\N$, and $\varepsilon^{(1)}, \varepsilon^{(2)}, \ldots, \varepsilon^{(d)} \sim 0$. Then there are $c_1, c_2, \ldots, c_d \in \K$ such that
\[
\nested_{i=1}^d \left(a^{(i)} + \varepsilon^{(i)} \right)\ \sim\ \sum_{i=1}^d c_i\, \nested_{j=1}^i a^{(j)}.
\]
\end{lemma}

\proof
By induction on $d$.

If $d=0$ both sides are 0. Now assume that the assertion holds at some $d \ge 1$, and expand the left-hand side. In line 2 we use the induction hypothesis and compensate for replacing equivalence with equality by adding a sequence $\eta \sim 0$ in the appropriate place. In lines 3 and 4 we use Lemma \ref{conveq}.(i) resp.\ (ii). We denote the constant introduced by  Lemma \ref{conveq}.(ii) by $c_1$:
\begin{align*}
\left(\nested_{i=1}^{d+1} \left(a^{(i)} + \varepsilon^{(i)} \right)\right)_n &=\ \left(a^{(1)}_n + \varepsilon^{(1)}_n \right) \sum_{k_2=0}^n \left(\nested_{i=2}^{d+1} \left(a^{(i)} + \varepsilon^{(i)} \right)\right)_{k_2} \\
&=\ \left(a^{(1)}_n + \varepsilon^{(1)}_n \right) \sum_{k_2=0}^n \left(\sum_{i=2}^{d+1} c_i\, \left(\nested_{j=2}^i\, a^{(j)}\right)_{k_2} + \eta_{k_2}\right) \\
&\sim\ a^{(1)}_n \sum_{k_2=0}^n \sum_{i=2}^{d+1} c_i\, \left(\nested_{j=2}^i\, a^{(j)}\right)_{k_2} +  a^{(1)}_n \sum_{k_2=0}^n \eta_{k_2} \\
&\sim\ \sum_{i=2}^{d+1} c_i \, \left(\nested_{j=1}^i\, a^{(j)}\right)_n +  c_1 a^{(1)}_n \\
&=\ \sum_{i=1}^{d+1} c_i \, \left(\nested_{j=1}^i\, a^{(j)}\right)_n. 
\end{align*}
\qed

\begin{lemma}
\label{lemmaN}
Let $d \in \N$ and $a^{(1)}, a^{(2)}, \ldots, a^{(d)} \in \K^\N$. If $N \in \N$ is s.t.\ $a^{(i)}_n = 0$ for all $n < N$ and $i \in \{1,2,\ldots,d\}$, then
\[
E^N \left(\nested_{i=1}^d a^{(i)}\right)\ =\ \nested_{i=1}^d\, E^N \left(a^{(i)}\right).
\]
\end{lemma}

\proof
Write the nested sum on the left as a single sum, shift all summation indices by $N$, and use the fact that all original summands vanish below $N$:
\begin{eqnarray*}
\left(E^N \left(\nested_{i=1}^d a^{(i)}\right)\right)_n &=&  a^{(1)}_{n+N} \sum_{k_2=0}^{n+N} a^{(2)}_{k_2} \sum_{k_3=0}^{k_2} a^{(3)}_{k_3} \cdots \sum_{k_{d}=0}^{k_{d-1}} a^{(d)}_{k_d} \\
&=& \sum_{0\le k_d\le \cdots\le k_3 \le k_2\le n+N} a^{(1)}_{n+N} a^{(2)}_{k_2} a^{(3)}_{k_3} \cdots a^{(d)}_{k_d} \\
&=& \sum_{-N\le k_d\le \cdots\le k_3 \le k_2\le n} a^{(1)}_{n+N} a^{(2)}_{k_2+N} a^{(3)}_{k_3+N} \cdots a^{(d)}_{k_d+N} \\
&=& \sum_{0\le k_d\le \cdots\le k_3 \le k_2\le n} a^{(1)}_{n+N} a^{(2)}_{k_2+N} a^{(3)}_{k_3+N} \cdots a^{(d)}_{k_d+N} \\
&=& \left(\nested_{i=1}^d\, E^N \left(a^{(i)}\right)\right)_n.
\end{eqnarray*}
\qed

\begin{theorem}
\label{main}
Let $d \in \N$, $\eta_1, \eta_2, \ldots, \eta_d \in \N$, and $h^{(1)}, h^{(2)}, \ldots, h^{(d)} \in {\cal H}(\K)$. Let $p_1, p_2, \ldots, p_d \in \K[x]$, $q_1, q_2, \ldots, q_d \in \K[x]$ be such that $q_i(n)h^{(i)}_{n+1} = p_i(n)h^{(i)}_n$, $q_i(n) \ne 0$ for all $n \in \N$ and $i \in \{1,2,\ldots,d\}$. Let $e \in \N$,  $\xi_1, \xi_2, \ldots, \xi_e \in \N$, and $\varphi_i(x) \in \{\alpha_i^x x^{j_i}, \alpha_i^x (x - \beta_i)^{-j_i}\}$ for all $i  \in \{1,2,\ldots, e\}$  where $j_1, j_2, \ldots, j_e \in\N$, $\alpha_1, \alpha_2, \ldots, \alpha_e \in \K^*$ and $\beta_1, \beta_2, \ldots, \beta_e \in \K\setminus \N$. Let $a \in \K^\N$ be given by $a = 0$ if $d = 0$, and
\begin{align*}
a_n &= \displaystyle
h^{(1)}_{n} \sum_{k_2=0}^{n+\eta_1} h^{(2)}_{k_2} \sum_{k_3=0}^{k_2+\eta_2} h^{(3)}_{k_3}\ \cdots \!\!\sum_{k_{d}=0}^{k_{d-1}+\eta_{d-1}}  h^{(d)}_{k_d} \ \ {\rm for\ all\ } n \in \N
\end{align*}
if $d \ge 1$. Let $b \in \K^\N$ be given by $b = 0$ if $e = 0$, and
\begin{align*}
b_n &= \displaystyle
\varphi_1(n) \sum_{k_2=0}^{n+\xi_1} \varphi_2(k_2) \sum_{k_3=0}^{k_2+\xi_2} \varphi_3(k_3) \ \cdots \!\!\sum_{k_e=0}^{k_{e-1}+\xi_{e-1}} \varphi_e(k_e) \ \ {\rm for\ all\ } n \in \N
\end{align*}
if $e \ge 1$. Then $a * b$ is d'Alembertian.
\end{theorem}

\proof
By  induction on $d+e+\sum_{i=1}^e j_i$ where $d$ resp.\ $e$ are the nesting depths of these representations of $a$ resp.\ $b$, and $\sum_{i=1}^e j_i$ is the \emph{valuation} of $b$. If $d = 0$ or $e = 0$, then $a * b = 0 \in {\cal A}(\K)$. Now let $d, e \ge 1$. Write
\begin{equation}
\label{un}
a_n\ =\  h_n \sum_{k_2=0}^{n+\eta} \tilde a_{k_2}
\end{equation}
where $h = h^{(1)}$, $\eta = \eta_1$ and 
\[
\tilde a_{k_2}\ =\ \left\{
\begin{array}{ll}
\delta_{k_2,0}\ \dotfill, & d = 1, \\ \displaystyle
h^{(2)}_{k_2} \sum_{k_3=0}^{k_2+\eta_2} h^{(3)}_{k_3}\ \cdots \sum_{k_{d}=0}^{k_{d-1}+\eta_{d-1}}  h^{(d)}_{k_d}, & d \ge 2 
\end{array}
\right.
\]
with $\delta_{k_2,0} = \langle 1,0,0,0,\ldots \rangle \in \K^\N$ the identity element for convolution. Write
\begin{equation}
\label{vn}
b_n\ =\  \varphi(n) \sum_{k_2=0}^{n+\xi} \tilde b_{k_2}
\end{equation}
where $\varphi = \varphi_1$ (with $j = j_1$, $\alpha = \alpha_1$, $\beta = \beta_1$), $\xi=\xi_1$ and 
\[
\tilde b_{k_2}\ =\ \left\{
\begin{array}{ll}
\delta_{k_2,0}\ \dotfill, & e = 1, \\ \displaystyle
\varphi_2(k_2) \sum_{k_3=0}^{k_2+\xi_2} \varphi_3(k_3)\ \cdots \sum_{k_e=0}^{k_{e-1}+\xi_{e-1}} \varphi_e(k_e), & e \ge 2.
\end{array}
\right.
\]
We shall prove that the convolution
\[
y_n\ :=\ \sum_{k=0}^n a_k b_{n-k}\ =\ \sum_{k=0}^n h_k \varphi(n-k) \left(\sum_{k_2=0}^{k+\eta} \tilde a_{k_2}\right) \left(\sum_{k_2=0}^{n-k+\xi} \tilde b_{k_2}\right)
\]
is d'Alembertian by showing that $L_0 (y) \in {\cal A}(\K)$ for an appropriate operator $L_0 \in \K[n]\langle E\rangle$, 
then invoking Corollary \ref{dalembRHS}. We distinguish three cases:

\bigskip
\textsc{Case 1.} $\varphi(x)=\alpha^x$

\smallskip
\noindent
In this case $y_n = \sum_{k=0}^n a_k \alpha^{n-k}\sum_{k_2=0}^{n-k+\xi} \tilde b_{k_2}$ and we take $L_0 = E - \alpha$. Then
\begin{eqnarray*}
\lefteqn{(L_0(y))_n \ =\ y_{n+1} - \alpha y_n}\\
 &=& \sum_{k=0}^{n+1} a_k \alpha^{n+1-k}\sum_{k_2=0}^{n+1-k+\xi} \tilde b_{k_2} - \sum_{k=0}^n a_k \alpha^{n+1-k} \sum_{k_2=0}^{n-k+\xi} \tilde b_{k_2} \\
 &=& a_{n+1} \sum_{k_2=0}^{\xi}\tilde b_{k_2} + \sum_{k=0}^n a_k \alpha^{n+1-k} \tilde b_{n-k+\xi+1}\\
&=&  E(a)_{n} \sum_{k_2=0}^{\xi}\tilde b_{k_2} + \alpha \left(a_n * \alpha^n E^{\xi+1}(\tilde b)_{n}\right)
\end{eqnarray*}
where $E(a)$ is d'Alembertian and $\alpha^n E^{\xi+1}(\tilde b)_{n}$ has nesting depth $e-1$, hence $a_n * \alpha^n E^{\xi+1}(\tilde b)_{n}$ is d'Alembertian by induction hypothesis.

\bigskip
\textsc{Case 2.} $\varphi(x) = \alpha^x x^j$ with $j \ge 1$

\medskip
\noindent
Here $b_n = n^j c_n$ where $c_n = \alpha^n \sum_{k_2=0}^{n+\xi} \tilde b_{k_2}$, and we take $L_0 = 1$. So 
\begin{align*}
y_n &=  \sum_{k=0}^n a_k (n-k)^j c_{n-k} = \sum_{i=0}^j (-1)^i \binom{j}{i} n^{j-i} \sum_{k=0}^n k^i a_k c_{n-k} \\
&= \sum_{i=0}^j (-1)^i \binom{j}{i} n^{j-i} \left((n^i a_n) * c_n \right).
\end{align*}
As the valuation of $c$ is $j$ less than that of $b$, our induction hypothesis implies that $y$ is d'Alembertian.

\medskip
\textsc{Case 3.} $\varphi(x) = \frac{\alpha^x}{(x-\beta)^{j}}$ with $j \ge 1$

\smallskip
\noindent
Here we take $L_0 = q(n-\beta) E - p(n-\beta)$ where polynomials $p, q \in \K[n] \setminus\{0\}$ are such that $q(n) h_{n+1} - p(n)h_n = 0$ for all $n \in \N$.  Then
\begin{eqnarray*}
\lefteqn{(L_0 (y))_n\ = } \\
&=& q(n-\beta) \sum_{k=0}^{n+1} \frac{h_k \alpha^{n+1-k}}{(n+1 - k - \beta)^j} \left(\sum_{k_2=0}^{k+\eta} \tilde a_{k_2}\right) \left(\sum_{k_2=0}^{n+1-k+\xi} \tilde b_{k_2}\right)\\
&-& p(n-\beta)\sum_{k=0}^n \frac{h_k \alpha^{n-k}}{(n-k-\beta)^j}\left(\sum_{k_2=0}^{k+\eta} \tilde a_{k_2}\right) \left(\sum_{k_2=0}^{n-k+\xi} \tilde b_{k_2}\right) \\
&=& q(n-\beta) \sum_{k=-1}^{n} \frac{h_{k+1}\alpha^{n-k}}{(n - k - \beta)^j} \left(\sum_{k_2=0}^{k+1+\eta} \tilde a_{k_2}\right) \left(\sum_{k_2=0}^{n-k+\xi} \tilde b_{k_2}\right)\\
&-& p(n-\beta)\sum_{k=0}^n \frac{h_k \alpha^{n-k}}{(n-k-\beta)^j}\left(\sum_{k_2=0}^{k+\eta} \tilde a_{k_2}\right) \left(\sum_{k_2=0}^{n-k+\xi} \tilde b_{k_2}\right) \\
&=& A_n + B_n + C_n,
\end{eqnarray*}
where
\begin{eqnarray*}
A_n &:=& q(n-\beta) a_0 \, b_{n+1}, \\
B_n &:=& q(n-\beta) \sum_{k=0}^n h_{k+1} \tilde a_{k+1+\eta} \frac{\alpha^{n-k}}{(n-k-\beta)^j}\left(\sum_{k_2=0}^{n-k+\xi} \tilde b_{k_2}\right) , \\
C_n &:=&  \sum_{k=0}^{n} \frac{q(n-\beta) h_{k+1} - p(n-\beta)h_k}{(n - k - \beta)^j}\alpha^{n-k} \left(\sum_{k_2=0}^{k+\eta} \tilde a_{k_2}\right) \left(\sum_{k_2=0}^{n-k+\xi} \tilde b_{k_2}\right).
\end{eqnarray*}
Clearly $A$ is d'Alembertian. Since $B_n = q(n-\beta)\left( E(h E^\eta(\tilde a)) * b\right)_n$ and the nesting depth of $E(h E^\eta(\tilde a))$ is $d-1$, $B$ is d'Alembertian by the induction hypothesis. In $C_n$ we replace $h_{k+1}$ with $h_k\, p(k)/q(k)$ and obtain
\begin{eqnarray*}
\label{nkbj}
C_n &=&  \sum_{k=0}^n \frac{P(k) h_k \alpha^{n-k}}{q(k)(n - k - \beta)^j}  \left(\sum_{k_2=0}^{k+\eta} \tilde a_{k_2}\right) \left(\sum_{k_2=0}^{n-k+\xi} \tilde b_{k_2}\right)
\end{eqnarray*}
where $P(k) := q(n-\beta)p(k)- p(n-\beta)q(k) \in \K[n][k]$. Since $P(n-\beta) = 0$, $P(k)$ is divisible by $k-n+\beta$, hence there are $s\in\N$ and $c_0, c_1, \ldots, c_s \in \K[x]$ such that $P(k) = (n-k-\beta)\sum_{i=0}^s c_i(n)k^i$. It follows that
\begin{eqnarray*}
C_n &=&  \sum_{i=0}^s c_i(n) \sum_{k=0}^{n}  \frac{ k^i  a_k}{q(k)}\cdot \frac{\alpha^{n-k}}{(n - k - \beta)^{j-1}} \left(\sum_{k_2=0}^{n-k+\xi} \tilde b_{k_2}\right)\\
&=&  \sum_{i=0}^s c_i(n) \sum_{k=0}^{n}  u_k^{(i)}\cdot \frac{\alpha^{n-k}}{(n - k - \beta)^{j-1}} \left(\sum_{k_2=0}^{n-k+\xi} \tilde b_{k_2}\right) \\
&=& \sum_{i=0}^s c_i(n) \left(u_n^{(i)} * (n-\beta) b_n\right)
\end{eqnarray*}
where $u_k^{(i)} :=  \frac{ k^i  a_k}{q(k)}$ for all $k \in \N$ and $i \in \{1, 2, \ldots, s\}$. As the valuation of $(n-\beta) b_n$ is one less than that of $b_n$, our induction hypothesis implies that $u_n^{(i)} * (n-\beta) b_n$ is d'Alembertian, hence so are $C$ and $L_0 (y) = A+B+C$. Since $L_0$ has order one, Corollary \ref{dalembRHS} implies that $y = a * b$ is d'Alembertian.  
\qed

\begin{corollary}
\label{dalemb*ratdalemb}
If $a \in \K^\N$ is d'Alembertian and $b \in \K^\N$ is (quasi-)rationally d'Alembertian, then their convolution $a * b$ is d'Alembertian.
\end{corollary}

\proof
By Theorem \ref{dalembprop}.(i), the sequence $a$ can be written as a $\K$-linear combination of sequences of the form $\nested_{i=1}^d h^{(i)}$ where $h^{(1)}, h^{(2)}, \ldots, h^{(d)} \in {\cal H}(\K)$. For $i = 1,2,\ldots, d$, let $p_i, q_i \in \K[x]$ be such that $q_i(n)h^{(i)}_{n+1} = p_i(n) h^{(i)}_n$ for all $n \in \N$.
By Theorem \ref{ratdalembprop}, the sequence $b$ can be written as a $\K$-linear combination of sequences of the form $\nested_{i=1}^e r^{(i)}$ where $r^{(1)}, r^{(2)}, \ldots, r^{(e)}$ are (quasi-)rational sequences.
By the Partial Fraction Decomposition Theorem for rational functions, we can assume that for $i = 1,2,\ldots, e$ there are $j_i \in \N$, $\alpha_i \in \K*$ and $\beta_i \in \K$ such that $r^{(i)}_n = \varphi_i(n)$ for all large enough $n$, where $\varphi_i(x) \in \{\alpha_i^x x^{j_i}, \alpha_i^x (x-\beta_i)^{-j_i}\}$.

 By bilinearity of convolution, it suffices to prove that the convolution of a single $\nested_{i=1}^d h^{(i)}$ with a single $\nested_{i=1}^e r^{(i)}$ is d'Alembertian, so henceforth we assume that $a \equiv \nested_{i=1}^d h^{(i)}$ and $b \equiv \nested_{i=1}^e r^{(i)}$. Let $N \in \N$ be such that $q_i(n) \ne 0$ and $r^{(i)}_n = \varphi_i(n)$ for all $n \ge N$ and $i \in \{1,2,\ldots,e\}$. We shall prove by induction on the sum of nesting depths $d+e$ that $a * b$ is d'Alembertian.

If $d=0$ or $e=0$ then $a=0$ or $b=0$ and so $a*b = 0 \in {\cal A}(\K)$.

Assume now that $d \ge 1$ and $e \ge 1$. Let $\tilde a := \nested_{k=1}^d \tilde h^{(k)}$ and $\tilde b := \nested_{k=1}^e \tilde r^{(k)}$ where $\tilde h^{(k)} = E_0^{-N}E^N\left(h^{(k)}\right)$ and $\tilde r^{(k)} = E_0^{-N}E^N\left(r^{(k)}\right)$.
Then $\tilde h^{(k)}_n = \tilde r^{(k)}_n = 0$ for $n < N$ and $\tilde h^{(k)}_n =  h^{(k)}_n$, $\tilde r^{(k)}_n =  r^{(k)}_n$ for $n \ge N$. It follows by Lemma \ref{lemmaN} that
\begin{align*}
E^N(\tilde a)_n &= \nested_{k=1}^d E^N\left(\tilde h^{(k)}\right)_n =  \nested_{k=1}^d E^N\left(h^{(k)}\right)_n =  \nested_{k=1}^d h^{(k)}_{n+N} , \\
E^N(\tilde b)_n &= \nested_{k=1}^e E^N\left(\tilde r^{(k)}\right)_n =  \nested_{k=1}^e E^N\left(r^{(k)}\right)_n = \nested_{k=1}^e \varphi_k(n+N).
\end{align*}
Note that by our definition of $N$, the sequences $E^N(\tilde a)$ and $E^N(\tilde b)$ satisfy all the assumptions of Theorem \ref{main}, so $E^N(\tilde a) * E^N(\tilde b) \in {\cal A}(\K)$. Proposition \ref{dalembEa*Eb} now implies that $\tilde a * \tilde b \in {\cal A}(\K)$ as well.

By Lemma \ref{lemmaZ}, there are $c_1, c_2, \ldots, c_d \in \K$ and $c'_1, c'_2, \ldots, c'_e \in \K$ such that
\begin{align*}
a &= \sum_{i=1}^d c_i \,\nested_{j=1}^i \tilde h^{(j)} + \eta =  c_d \,\nested_{j=1}^d \tilde h^{(j)} + \sum_{i=1}^{d-1} c_i \,\nested_{j=1}^i \tilde h^{(j)} + \eta, \\
b &= \sum_{i=1}^e c'_i \,\nested_{j=1}^i \tilde r^{(j)} + \eta' =  c'_e \,\nested_{j=1}^e \tilde r^{(j)} + \sum_{i=1}^{e-1} c'_i \,\nested_{j=1}^i \tilde r^{(j)} + \eta'
\end{align*}
for some sequences $\eta, \eta' \sim 0$. Hence
\begin{align*}
a * b\ &=\  c_d \,c'_e \,\nested_{j=1}^d \tilde h^{(j)} * \nested_{j=1}^e \tilde r^{(j)}  +  c_d \,\nested_{j=1}^d \tilde h^{(j)} *  \sum_{i=1}^{e-1} c'_i \,\nested_{j=1}^i \tilde r^{(j)}  \\
&+\  c'_e \,\nested_{j=1}^e \tilde r^{(j)} * \sum_{i=1}^{d-1} c_i \,\nested_{j=1}^i \tilde h^{(j)} + \sum_{i=1}^{d-1} c_i \,\nested_{j=1}^i \tilde h^{(j)} *   \sum_{i=1}^{e-1} c'_i \,\nested_{j=1}^i \tilde r^{(j)} \\
&+\ \eta * \left(c'_e \,\nested_{j=1}^e \tilde r^{(j)} + \sum_{i=1}^{e-1} c'_i \,\nested_{j=1}^i \tilde r^{(j)} + \eta'\right) \\
&+\ \eta' * \left( c_d \,\nested_{j=1}^d \tilde h^{(j)} + \sum_{i=1}^{d-1} c_i \,\nested_{j=1}^i \tilde h^{(j)}\right).
\end{align*}
The first term on the right equals $ c_d \,c'_e \, \tilde a * \tilde b$, so it is d'Alembertian as shown in the previous paragraph. The next three terms are linear combinations of convolutions of nested sums having nesting depths at most $d+e-1$,  $d+e-1$, and  $d+e-2$, respectively, so they are d'Alembertian by induction hypothesis. By Lemma \ref{conveq}.(iii), the last two terms above are linear combinations of shifted d'Alembertian sequences, so they are d'Alembertian as well. It follows that $a * b$ is d'Alembertian as claimed. \qed

\begin{example}
\label{a0*f0}
By Corollary \ref{dalemb*ratdalemb}, the convolution of a hypergeometric sequence with a rational sequence, such as
\[
y_n\ =\ (2^{n-1}n!)*\left(\cfrac{1}{n+\frac{1}{2}}\right)\ =\ \sum_{k=0}^n \frac{2^{k-1}k!}{n-k+\frac{1}{2}},
\]
is d'Alembertian. By following through the proof of Theorem \ref{main} with $a_n = 2^{n-1}n!$ and $b_n = 1/(n+\frac{1}{2})$, we will obtain an explicit nested-sum representation of $y_n$. Here the nesting depths of $a$ and $b$ are $1$, $j = 1$, $\beta = -1/2$, $h_n = a_n = 2^{n-1}n!$, $h_{n+1}/h_n = p(n) = 2(n+1)$, $q(n)=1$ and 
\[
L_0\ =\ q(n-\beta) E - p(n-\beta)\ =\ E-(2n+3).
\]
Applying $L_0$ to $y(n)$ we obtain
\begin{eqnarray}
(L_0(y))_n &=& \sum_{k=0}^{n+1} \frac{2^{k-1}k!}{n-k+\frac{3}{2}} - (2n+3)\sum_{k=0}^n \frac{2^{k-1}k!}{n-k+\frac{1}{2}}\nonumber \\ \nonumber
&=& \sum_{k=-1}^{n} \frac{2^{k}(k+1)!}{n-k+\frac{1}{2}} - (2n+3)\sum_{k=0}^n \frac{2^{k-1}k!}{n-k+\frac{1}{2}} \\ \nonumber
&=& \frac{1}{2n+3} + \sum_{k=0}^n \frac{2^{k-1}k!(2(k+1)-(2n+3))}{n-k+\frac{1}{2}} \\ \nonumber
&=& \frac{1}{2n+3} + \sum_{k=0}^n \frac{2^{k-1}k!(2k-2n-1)}{n-k+\frac{1}{2}} \\ 
&=& \frac{1}{2n+3} - \sum_{k=0}^n 2^{k}k! \ =\  \frac{1}{2n+3} - \sum_{k=0}^n (2k)!!. \label{indefsum}
\end{eqnarray}
By solving this recurrence with initial condition $y_0 = 1$, we obtain
\[
y_n\ =\ (2n+1)!!\left(1+\sum_{k=1}^n \frac{1}{(2k+1)!!}\left(\frac{1}{2k+1}-\sum_{j=0}^{k-1} (2j)!!\right)\right).
\]
Since $(2n+1)!!$ and $(2n)!!$ are hypergeometric sequences, this shows that $y(n)$ is indeed a d'Alembertian sequence. 

From {\em (\ref{indefsum})} we can also obtain a fully factored annihilator of $y$ as follows: The right-hand side of  {\em (\ref{indefsum})} is annihilated by the least common left multiple of $E - (2 n + 3)/(2 n + 5)$ and $(E - (2n+4))(E - 1)$, which is
\[
\left(E - \frac{(2n+3)(2n+7)^2}{(2n+5)^2(2n+9)}\right)\left(E - (2n+4)\right)(E - 1),
\]
hence $L (y) = 0$ where
\[
L\ =\ \left(E - \frac{(2n+3)(2n+7)^2}{(2n+5)^2(2n+9)}\right)\left(E - (2n+4)\right)(E - 1)(E-(2n+3)).
\]
\end{example}

\begin{example}
The sequence
\[
y_n = H_n * H_n = \sum_{k=0}^n H_k H_{n-k} = \sum_{k=0}^n \left(\sum_{i=1}^k \frac{1}{i} \sum_{j=1}^{n-k} \frac{1}{j}\right)
\]
 is annihilated by $L = \left((n + 3) E - (n + 2)\right)^2 (E - 1)^2$.
\end{example}

\begin{example}
The sequence
\[
y_n = n! * H_n = \sum_{k=0}^n k!\, H_{n-k} = \sum_{k=0}^n \left(k!\, \sum_{j=1}^{n-k} \frac{1}{j}\right)
\]
 is annihilated by $L = ((n + 5) E - (n + 4))\left((E  - (n + 2))(E  - 1)\right)^2$.
\end{example}

\section{Liouvillian sequences under convolution}
\label{secliouv}

Here we establish some connections between convolution, interlacing and inverse shift $E_0^{-1}$ which allow us to transfer results about convolutions of d'Alem\-ber\-ti\-an sequences to the corresponding results about Liouvillian sequences. Recall that the (ordinary) generating series of a sequence $a \in \K^\N$ is defined as the formal power series
\[
   g_a(x) = \sum_{k=0}^\infty a_k x^k,
\]
and that for all pairs of sequences $a, b \in \K^\N$,
\[
   g_{a + b}(x) = g_a(x) +  g_b(x), \quad  g_{a*b}(x) = g_a(x) g_b(x).
\]
\begin{definition}
{\em \cite{HS}}
For $m \in \N\setminus \{0\}$ and $a \in \K^\N$, we write $\Lambda^m a$ for $\Lambda(a,\overbrace{0,\ldots,0}^{m-1})$, and call it the \emph{$0^{\rm th}$ $m$-interlacing of $a$ with zeroes}.
\end{definition}

\begin{lemma}
\label{lem:aux}
Let $k \in \N$, $m \in \N \setminus \{0\}$, $a, a^{(0)}, a^{(1)}, \ldots, a^{(m-1)} \in \K^\N$, and $b =\Lambda_{j=0}^{m-1} a^{(j)}$.
\begin{enumerate}
\item[\rm (i)] $\left( E_0^{-k}(a)\right)_n = \left\{
\begin{array}{ll}
a_{n-k}, & n \ge k \\
0, & n < k
\end{array}
\right.$

\item[\rm (ii)] $\left(\Lambda^m a\right)_n\ =\ \left\{
\begin{array}{ll}
a_{\frac{n}{m}}, & n \equiv 0 \pmod{m} \\
0, & n \not\equiv 0 \pmod{m}
\end{array}
\right.$

\item[\rm (iii)] $g_{E_0^{-k}(a)}(x)\ =\ x^k g_a(x)$

\item[\rm (iv)] $g_{\Lambda^m a}(x)\ =\ g_a(x^m)$

\item[\rm (v)] $ g_b(x)\ =\ \sum_{j=0}^{m-1} x^j g_{a^{(j)}}(x^m)$

\item[\rm (vi)] $\Lambda_{j=0}^{m-1} a^{(j)}\ =\ \sum_{j=0}^{m-1} E_0^{-j} \left( \Lambda^m a^{(j)} \right)$

\item[\rm (vii)] $\Lambda^m E_0^{-k}\ =\ E_0^{-k m} \Lambda^m$
\end{enumerate}
\end{lemma}

\proof
Items (i), (ii) follow immediately from the definitions of $E_0^{-1}$ and $\Lambda^m$.
\begin{itemize}
\item[\rm (iii):] $g_{E_0^{-k}(a)}(x)\ =\ \displaystyle \sum_{n=0}^\infty E_0^{-k}(a)_{n} x^n\ =\ \sum_{n=k}^\infty a_{n-k} x^n\ =\ \sum_{n=0}^\infty a_{n} x^{n+k}\ =\ x^k g_a(x)$

\item[\rm (iv):] $g_{\Lambda^m a}(x)\ =\ \displaystyle \sum_{n=0}^\infty(\Lambda^m a)_{n} x^n\ =\ \sum_{n \equiv 0 \pmod{m}} a_{\frac{n}{m}} x^n\ =\ \sum_{k=0}^\infty a_k x^{k m}\ =\ g_a(x^m)$

\item[\rm (v):] 
$g_b(x) = \sum_{n=0}^\infty a_{n \dd m}^{(n \bmod m)} x^n = \sum_{j=0}^{m-1}\sum_{k=0}^\infty a_k^{(j)} x^{km+j} = \sum_{j=0}^{m-1} x^j g_{a^{(j)}} (x^m)$

\item[\rm (vi):] Using (v), (iv) and (iii) we find that
\begin{align*}
g_b(x)\ &=\ \sum_{j=0}^{m-1} x^j g_{a^{(j)}} (x^m)\ =\ \sum_{j=0}^{m-1} x^j g_{\Lambda^m a^{(j)}} (x)\ =\ \sum_{j=0}^{m-1} g_{ E_0^{-j} \left( \Lambda^m a^{(j)} \right)}(x) \\
&=\ g_{\sum_{j=0}^{m-1} E_0^{-j} \left( \Lambda^m a^{(j)} \right)}(x)
\end{align*}
which implies the assertion. 

\item[\rm (vii):] By applying (iv) and (iii) alternatingly, we obtain
\begin{align*}
\!\!\!\!\! g_{\Lambda^m E_0^{-k}(a)}(x) &= g_{E_0^{-k}(a)}(x^m) = x^{k m}g_a(x^m) = x^{k m}g_{\Lambda^m a}(x) = g_{E_0^{-k m} \Lambda^m a}(x)
\end{align*}
for every $a  \in \K^\N$, which implies the assertion. \qed
\end{itemize}

\begin{proposition}
\label{convd-zero}
The convolution of the $0^{\rm th}$ $m$-interlacings of $a, b \in \K^\N$ with zeroes is the $0^{\rm th}$ $m$-interlacing of $a*b$ with zeroes:
\[
   \Lambda^m a * \Lambda^m b = \Lambda^m (a*b).
\]
\end{proposition}

\proof
\begin{eqnarray*}
  g_{\Lambda^m a * \Lambda^m b}(x) &=& g_{\Lambda^m a}(x) g_{\Lambda^m b}(x)= g_a(x^m) g_b(x^m)\\
    &=& \sum_{i=0}^\infty a_i x^{mi} \sum_{j=0}^\infty b_j x^{mj}= \sum_{i=0}^\infty \sum_{j=0}^\infty a_i b_j x^{m(i+j)}\\
    &=& \sum_{k=0}^\infty x^{mk} \sum_{i=0}^k a_i b_{k-i}= \sum_{k=0}^\infty (a*b)_k  \left(x^m\right)^k \\
    &=&  g_{a*b} (x^m) = g_{\Lambda^m (a*b)} (x)
\end{eqnarray*}
by using Lemma \ref{lem:aux}.(iv) three times.
\qed

\begin{proposition}
\label{d-u*v}
Let $m \in \N \setminus \{0\}$, $a^{(j)}, b^{(j)} \in \K^\N$ for all $j \in \{0,1,\ldots,m-1\}$, $u =\Lambda_{j=0}^{m-1} a^{(j)}$, and $v =\Lambda_{j=0}^{m-1} b^{(j)}$. Then
\begin{equation}  
\label{eq-d-u*v}
   u*v = \sum_{k=0}^{2m-2} \sum_{j=\max\{0,k-m+1\}}^{\min\{k,m-1\}}  E_0^{-k} \Lambda^m\left(a^{(j)}*b^{(k-j)}\right).
\end{equation}   
\end{proposition}

\proof
Using Lemma \ref{lem:aux}.(v),  we obtain
\begin{eqnarray}
  g_{u*v}(x) &=&  g_u(x) g_v(x)= \sum_{j=0}^{m-1} x^j g_{a^{(j)}}(x^m) \sum_{\ell=0}^{m-1} x^\ell g_{b^{(\ell)}}(x^m)\nonumber\\
    &=&  \sum_{j=0}^{m-1} \sum_{\ell=0}^{m-1} x^{j+\ell} g_{a^{(j)}}(x^m) g_{b^{(\ell)}}(x^m)\nonumber\\
    &=& \sum_{k=0}^{m-1}  \sum_{j=0}^{k} + \sum_{k=m}^{2m-2} \sum_{j=k-m+1}^{m-1} x^k g_{a^{(j)}}(x^m)g_{b^{(k-j)}}(x^m) \nonumber\\
    &=& \sum_{k=0}^{2m-2} \sum_{j=\max\{0,k-m+1\}}^{\min\{k,m-1\}} x^k g_{a^{(j)}}(x^m)g_{b^{(k-j)}}(x^m) \label{long}
\end{eqnarray}
By Lemma \ref{lem:aux}.(iv), Proposition \ref{convd-zero} and  Lemma \ref{lem:aux}.(iii),
\begin{align*}
 x^k g_{a^{(j)}}(x^m)g_{b^{(k-j)}}(x^m) &= x^k g_{\Lambda^m a^{(j)}}(x)g_{\Lambda^m b^{(k-j)}}(x) =  x^k g_{\Lambda^m a^{(j)}*\Lambda^m b^{(k-j)} }(x)\\
&= x^k g_{\Lambda^m \left( a^{(j)}*b^{(k-j)} \right)}(x) =  g_{ E_0^{-k}\Lambda^m \left( a^{(j)}*b^{(k-j)} \right)}(x)
\end{align*}
which, together with (\ref{long}), implies (\ref{eq-d-u*v}).
\qed

\begin{corollary}
\label{liou*rat}
The convolution of a Liouvillian sequence $u$ with a (quasi-)ra\-tio\-nal\-ly Liouvillian sequence $v$ is Liouvillian.
\end{corollary}

\proof
Let $u=\Lambda_{i=0}^{m-1} a^{(i)}$ with all $a^{(i)}$ d'Alembertian, and $v=\Lambda_{i=0}^{k-1} b^{(i)}$ with all $b^{(i)}$ (quasi-)rationally d'Alembertian. Let $\ell = {\rm lcm}(m,k)$. Write $u=\Lambda_{j=0}^{\ell-1} c^{(j)}$, $v=\Lambda_{j=0}^{\ell-1} d^{(j)}$ where $c^{(j)}$ and $d^{(j)}$, for $j=0, 1, \ldots, \ell-1$, are the $j$-th $l$-sections of $u$ and $v$, respectively. Clearly all $c^{(j)}, d^{(j)}$ are  themselves sections of $a^{(i)}$ resp.\ $b^{(i)}$. Since ${\cal A}(\K)$ is closed under multisection \cite[Prop.\,7]{PZ13}, all $c^{(j)}$ are d'Alembertian. Similarly one can show that the ring of (quasi-)rationally d'Alembertian sequences is closed under multisection, hence all  $d^{(j)}$ are (quasi-)rationally d'Alem\-ber\-tian. So by Corollary \ref{dalemb*ratdalemb}, all convolutions $c^{(j_1)} * d^{(j_2)}$ for $j_1, j_2 \in \{0, 1, \ldots, \ell-1\}$ are d'Alembertian. It follows from Proposition \ref{d-u*v} that $u * v$ is a sum of shifted interlacings of d'Alembertian sequences, hence it is Liouvillian. \qed
 
\begin{example}
\label{n!!}
By Corollary  \ref{liou*rat}, the convolution of a Liouvillian sequence with a rational sequence, such as
\[
y_n\ :=\ n!!*\left(\frac{1}{n+1}\right)\ =\ \sum_{k=0}^n \frac{k!!}{n-k+1},
\]
is Liouvillian. By following the proof of Proposition \ref{d-u*v} with $u_n = n!!$ and $v_n = \frac{1}{n+1}$, we will obtain a representation of $y_n$ as an interlacing of d'Alembertian sequences. Here $m=2$, $u = \Lambda(a^{(0)}, a^{(1)})$ and $v =  \Lambda(b^{(0)}, b^{(1)})$, where
\begin{eqnarray*}
a^{(0)}_n &=& (2n)!! \ =\ 2^n n!, \\
a^{(1)}_n &=& (2n+1)!! \ =\ \frac{(2n+1)!}{2^n n!}, \\
b^{(0)}_n &=& v_{2n} \ =\ \frac{1}{2n+1}, \\
b^{(1)}_n &=& v_{2n+1} \ =\ \frac{1}{2n+2}. 
\end{eqnarray*}
By Proposition \ref{d-u*v} at $m = 2$,
\begin{eqnarray*}
   u*v &=& \Lambda^2 \left(a^{(0)}*b^{(0)}\right) + E_0^{-1} \Lambda^2  \left(a^{(0)}*b^{(1)} + a^{(1)}*b^{(0)}\right) + E_0^{-2} \Lambda^2 \left(a^{(1)}*b^{(1)}\right).
\end{eqnarray*}   
Denote
\begin{eqnarray*}
g^{(0)} &:=& a^{(0)} * b^{(0)} + E^{-1}\left(a^{(1)} * b^{(1)}\right), \\
g^{(1)} &:=& a^{(0)} * b^{(1)} + a^{(1)} * b^{(0)}.
\end{eqnarray*}
For any $a,b,c,d \in \K^\N$ we have $\Lambda(a+b, c+d) = \Lambda(a,c) + \Lambda(b,d)$, therefore
\[
\Lambda\left(g^{(0)},\, g^{(1)}\right)  = \Lambda\left(a^{(0)} * b^{(0)}, a^{(0)} * b^{(1)}\right) +
\Lambda\left(E_0^{-1}\left(a^{(1)} * b^{(1)}\right), a^{(1)} * b^{(0)}\right),
\]
which by Lemma \ref{lem:aux}.(vi) at $m=2$ equals
\begin{eqnarray*}
  \Lambda^2 \left(a^{(0)}*b^{(0)}\right) + E_0^{-1} \Lambda^2 \left(a^{(0)}*b^{(1)}+ a^{(1)}*b^{(0)}\right)+ \Lambda^2  E_0^{-1}\left(a^{(1)}*b^{(1)}\right).   
\end{eqnarray*}   
Since $\Lambda^2  E_0^{-1}\left(a^{(1)}*b^{(1)}\right) =  E_0^{-2} \Lambda^2 \left(a^{(1)}*b^{(1)}\right)$ by Lemma \ref{lem:aux}.(vii) at $m = 2$, it follows that $  u*v \ =\ \Lambda\left(g^{(0)},\, g^{(1)}\right)$.  It remains to show that $g^{(0)}$ and $g^{(1)}$ are d'Alembertian. We have
\begin{eqnarray*}
 (a^{(0)} * b^{(0)})_n &=& \sum_{k=0}^n \frac{2^k k!}{2(n-k)+1}\ =\ \sum_{k=0}^n \frac{2^{k-1} k!}{n-k+\frac{1}{2}}, \\
 (a^{(1)} * b^{(1)})_{n-1} &=& \sum_{k=0}^n \frac{(2k+1)!}{2^k k! (2(n-k-1)+2)}\ =\ \sum_{k=0}^n \frac{(2k+1)!}{2^{k+1} k! (n-k)}, \\
 (a^{(0)} * b^{(1)})_n &=& \sum_{k=0}^n \frac{2^k k!}{2(n-k)+2}\ =\ \sum_{k=0}^n \frac{2^{k-1} k!}{n-k+1}, \\
 (a^{(1)} * b^{(0)})_n &=& \sum_{k=0}^n \frac{(2k+1)!}{2^k k! (2(n-k)+1)}\ =\ \sum_{k=0}^n \frac{(2k+1)!}{2^{k+1} k! (n-k+\frac{1}{2})}.
\end{eqnarray*}
In an analogous way as we did it for $ (a^{(0)} * b^{(0)})_n$ in Example \ref{a0*f0}, we can compute explicit d'Alembertian representations  for $(a^{(1)} * b^{(1)})_{n-1}$, $(a^{(0)} * b^{(1)})_n$, and $(a^{(1)} * b^{(0)})_n$. After some additional simplification we obtain
\begin{eqnarray*}
g^{(0)}_n &=& (2n+1)!!\left(1+\sum_{k=1}^n \frac{1}{(2k+1)!!}\left(\frac{4k+1}{2k(2k+1)} - \sum_{j=0}^{2k-2} j!!\right)\right),\\
g^{(1)}_n &=& (2n+2)!!\left(\frac{3}{4}+\sum_{k=1}^n \frac{1}{(2k+2)!!}\left(\frac{4k+3}{(2k+1)(2k+2)} - \sum_{j=0}^{2k-1} j!!\right)\right).
\end{eqnarray*}
Note that
\begin{eqnarray*}
\sum_{j=0}^{2k-2} j!! &=&  \sum_{j=0}^{k-1}(2j)!! + \sum_{j=0}^{k-2}(2j+1)!!, \\
\sum_{j=0}^{2k-1} j!! &=&  \sum_{j=0}^{k-1}(2j)!! + \sum_{j=0}^{k-1}(2j+1)!!,
\end{eqnarray*}
hence both $g^{(0)}$ and $g^{(1)}$ are d'Alembertian sequences, and $u*v$, as their interlacing, is Liouvillian.
\end{example}

\section{Algorithmic considerations and some open problems}
\label{alg}

As we have seen, the ring of Liouvillian sequences is not closed under convolution. Several questions now arise naturally, such as:

\medskip\noindent
\textsc{Question:}
\textsf{Given a homogeneous linear recurrence equation with polynomial coefficients, how can we find all its solutions having the form of:
\begin{itemize}
\item a convolution of hypergeometric sequences,
\item a convolution of d'Alembertian sequences,
\item a convolution of Liouvillian sequences,
\item an expression built from hypergeometric sequences using the ten operations listed in Theorem \ref{ops}?
\end{itemize}
}

We do not have the answers to these questions (listed by increasing degree of difficulty), so we consider their relaxed versions where one of the factors of the convolutional solution is part of the input data for the problem. For example, the relaxed version of the simplest question above is the following:

\medskip\noindent
\textsc{Question:}
\textsf{Given a homogeneous linear recurrence equation with polynomial coefficients and a hypergeometric sequence $a$, how can we find all hypergeometric sequences $b$ such that $a * b$ is a solution of the given equation?}

\medskip
Even to this simple-looking question we do not have a complete answer. We do have a partial answer in the special case when the given sequence $a$ is \emph{hyperexponential}, i.e., its generating series $g_a(x) = \sum_{n=0}^\infty a_n x^n \in \K[[x]]$ satisfies the first-order linear differential equation
\[
g_a'(x) = r(x) g_a(x)
\]
for some rational function $r \in \K(x)$. Here we consider the ring of two-way sequences $\K^\Z$ on which the shift operator $E$ is an automorphism, and the action of operators from $\K[n]\langle E,E^{-1}\rangle$ on it. To each $a \in \K^\N$ we assign its \emph{padding with zeroes} $\zeta(a) \in \K^\Z$ defined by
\[
\zeta(a)_n = \left\{
\begin{array}{ll}
a_n, & n \ge 0, \\
0, & n < 0.
\end{array}\right.
\]
 We present here an algorithm which, given $L\in \K[n]\langle E\rangle $ and a fixed hyperexponential sequence $a$, returns an operator $L'\in \K[n]\langle E, E^{-1}\rangle $ such that $L' (\zeta(b)) = 0$ for all $b \in \K^\N$ with $L(\zeta(a * b)) = 0$. This algorithm is based on the well-known fact that $a \in \K^{\N}$ is P-recursive iff $\gf_a(x)$ is D-finite (see \cite[Thm.\ 1.4]{Stanley80}). Constructively, there exists a skew-Laurent-polynomial algebra isomorphism ${\cal R}:~\K[n]\langle E,E^{-1}\rangle \to \K[x,x^{-1}]\langle D\rangle$ (where $D = \frac{\partial}{\partial x}$ is the \emph{derivative operator}, $D g_a(x) = g_a'(x)$) which fixes each element of $\K$ and satisfies
\begin{equation}
\begin{array}{clll}
{\cal R}:& n &\mapsto & x D, \label{tr1} \\
{\cal R}:& E &\mapsto & x^{-1}, \\
{\cal R}:& E^{-1} &\mapsto & x,  
\end{array}
\end{equation}
\begin{equation}
\begin{array}{clll}
{\cal R}^{-1}:& x &\mapsto & E^{-1}, \label{tr3} \\
{\cal R}^{-1}:& x^{-1} &\mapsto & E, \\
{\cal R}^{-1}:& D &\mapsto & (n+1) E  
\end{array}
\end{equation}
(cf.\ \cite[Sec.\ 5]{APR}), such that for any $L \in \K[n]\langle E,E^{-1}\rangle $ and any $a \in \K^{\N}$ we have
\begin{equation}
\label{PRecDFin}
 L(\zeta(a)) = 0 \quad\Longleftrightarrow\quad  {\cal R}(L)(\gf_{a}(x)) = 0.
\end{equation}
The algorithm is as follows:

\bigskip
\begin{center}
\texttt{Algorithm HyperExpFactor}
\end{center}
\medskip\noindent
\verb|Input:  |$L \in \K[n]\langle E\rangle$ with $\ord L \ge 2$,\ \ $r \in \K(x)^*$
\vskip 7pt
\noindent
\verb|Output: |$L' \in \K[n]\langle E, E^{-1}\rangle$ \\

\noindent
\verb|  |1.\ Compute $M = {\cal R}(L) \in \K[x,x^{-1}]\langle D\rangle $  using (\ref{tr1}). \\

\noindent
\verb|  |2.\ Compute $M' \in \K[x,x^{-1}]\langle D\rangle $ such that  
\[
M'(v) = 0 \quad\Longleftrightarrow\quad M(u v) = 0
\]
\verb|    |whenever $u'(x) = r(x)u(x)$. \\

\noindent
\verb|  |3.\ Compute and return $L' = {\cal R}^{-1}(M') \in \K[n]\langle E,E^{-1}\rangle$  using (\ref{tr3}).

\begin{proposition}
Let  the output of algorithm \texttt{\em HyperExpFactor} with input $L, r$ be $L'$, and let $a \in \K^\N$ be s.t.\ $g_a'(x) = r(x) g_a(x)$. Then for every $b \in \K^\N$, we have
\[
L(\zeta(a * b)) = 0 \ \Longleftrightarrow \ L'(\zeta(b)) = 0.
\]
\end{proposition}

\proof
Going backwards through algorithm \texttt{HyperExpFactor}, we obtain
\begin{eqnarray*}
L'(\zeta(b)) = 0 
 &\Longleftrightarrow& {\cal R}^{-1}(M')(\zeta(b)) = 0 \\
 &\stackrel{{\rm by\,}(\ref{PRecDFin}){\rm\, with\, }L:= {\cal R}^{-1}(M')}{\Longleftrightarrow}&  M'(\gf_b(x)) = 0 \\
 &\Longleftrightarrow& M(\gf_a(x)\gf_b(x)) = 0\\
 &\Longleftrightarrow& {\cal R}(L)(\gf_{a*b}(x)) = 0 \\
 &\stackrel{{\rm by\,}(\ref{PRecDFin})}{\Longleftrightarrow}&  L(\zeta(a * b)) = 0 
\end{eqnarray*}
\qed

\begin{example}
Let $L$ be the operator from Example \ref{nonL} and let $a_n = 1/n!$, so that $r(x) = \gf_a'(x)/\gf_a(x) = e^x/e^x = 1$.
Running the above algorithm with input $L =  (n+3) E^3 - ( n^2 + 6 n + 10) E^2 + (2 n+5) E - 1$ and $r(x) = 1$ we obtain
\begin{enumerate}
\item $M = -x^{-2}(x^2 D^2  - (x-1) (2x-1) D + (x-2) (x-1))$,
\item $M' = -x^{-2}(x^2 D^2 + (3x-1) D + 1)$,
\item $L' = E^2((n+1)E - (n+1)^2)\ =\ (n+3)E^3 - (n+3)^2 E^2$.
\end{enumerate}
Denote $y = \zeta(b)$. Then
\[
(L'y)_n = (n+3) y_{n+3} - (n+3)^2 y_{n+2} = 0
\]
holds for all $n \le -3$ since $y_n = 0$ for $n < 0$. For $n \ge -2$ we can cancel the common factor $n+3$ and obtain $y_{n+3} - (n+3) y_{n+2} = 0$, or equivalently,
\[
y_{n+1} - (n+1) y_n = b_{n+1} - (n+1) b_n = 0
\]
for all $n \ge 0$. It follows that $b_n = C n!$ for $C \in \K$, in agreement with the fact that we constructed $L$ as an annihilator of $\frac{1}{n!} * n!$.
\end{example}

\section*{Acknowledgements}

S.A.A.\ acknowledges financial support from the Russian Foundation for Basic Research (project No.\ 16-01-001174). M.P.\ acknowledges financial support from the Slovenian Research Agency (research core funding No.\ P1-0294). The paper was completed while M.P.\ was attending the thematic programme ``Algorithmic and Enumerative Combinatorics'' at the Erwin Schr\"odinger International Institute for Mathematics and Physics in Vienna, Austria, to which he thanks for support and hospitality.


\begin{thebibliography}{9}

\bibitem{Abr91}
Abramov, S.A., 1991. An algorithm for finding quasi-rational solutions of differential and difference equations with polynomial coefficients (Russian). {\em Vestnik MGU} Ser.\ 15 (N 1), 43--48.

\bibitem{Abr93}
Abramov, S.A., 1993. On d'Alembert substitution. In: ISSAC '93. ACM, pp.\ 20--26.

\bibitem{AP94}
Abramov, S.A., Petkov\v sek, M., 1994. D'Alembertian solutions of linear operator equations. In:
ISSAC '94. ACM, pp.\ 169--174.

\bibitem{APR}
Abramov, S.A., Petkov\v sek, M., Ryabenko, A., 2000. Special formal series solutions of linear operator equations. {\em Discrete Math.} 210 (1--3), 3--25.

\bibitem{Gos78}
Gosper, R.W.\,Jr., 1978. Decision procedure for indefinite hypergeometric summation. 
{\em Proc.\ Nat.\ Acad.\ Sci.\ U.S.A.} 75 (1), 40--42.

\bibitem{HS}
Hendriks, P.A., Singer, M.F., 1999. Solving difference equations in finite terms.
{\em J.\ Symb.\ Comput.} 27 (3),  239--259.

\bibitem{Pet92}
Petkov\v sek, M., 1992. Hypergeometric solutions of linear recurrences with polynomial coefficients. {\em J.\ Symb.\ Comput.} 14 (2--3),  243--264.

\bibitem{Pet06}
Petkov\v sek, M., 2006. Symbolic computations over sequences (Russian).  {\em Programmirovanie} (2), 8--15; translation in {\em Program.\ Comput.\ Software} 32 (2), 65--70.

\bibitem{PZ13}
Petkov\v sek, M., Zakraj\v sek, H., 2013. Solving linear recurrence equations with polynomial coefficients. In: Schneider, C., Bl\"umlein, J., Eds.: {\em Computer Algebra in Quantum Field Theory: Integration, Summation and Special Functions}. Texts and Monographs in Symbolic Computation. Springer, Wien, pp.\ 259--284.

\bibitem{Reu12}
Reutenauer, C., 2012. On a matrix representation for polynomially recursive sequences. 
{\em Electron.\ J.\ Combin.} 19 (3), P36.

\bibitem{Stanley80}
Stanley,  R.P., 1980. Differentiably finite power series.
{\em European J.\ Combin.} 1 (2), 175--188.

\bibitem{Stanley99}
Stanley,  R.P., 1999.  \emph{Enumerative Combinatorics, Vol.\ 2}. 
Cambridge University Press, Cambridge.

\bibitem{Zeil90} 
Zeilberger, D., 1990. A fast algorithm for proving terminating hypergeometric identities.
{\em Discrete Math.} 80 (2), 207--211.

\bibitem{Zeil91} 
Zeilberger, D., 1991. The method of creative telescoping.
{\em J.\ Symb.\ Comput.} 11 (3), 195--204.

\end{thebibliography}
\end{document}